\documentclass[pdflatex,sn-mathphys-num]{sn-jnl}%

\usepackage{amsmath,amssymb,amsfonts}%
\usepackage{ulem}
\usepackage{todonotes}
\usepackage{siunitx}

\begin{document}

\title{Hyperuniformity as a unifying organizational principle across diatom architectures}

\author*[1]{\fnm{Eric} \sur{Ballestero}}\email{eric.ballestero@polito.it}
\author[2]{\fnm{Chiara} \sur{Gazzola}}
\author[3]{\fnm{Raj Kumar} \sur{Pal}}
\author[4]{\fnm{Vicent} \sur{Romero-Garc\'ia}}
\author*[1]{\fnm{Marco} \sur{Miniaci}}\email{marco.miniaci@gmail.com}

\affil[1]{\orgdiv{Department of Structural, Geotechnical and Building Engineering}, \orgname{Politecnico di Torino}, \orgaddress{\street{Corso Duca degli Abruzzi, 24}, \city{Torino}, \postcode{10129}, \country{Italy}}}

\affil[2]{\orgdiv{Department of Civil and Environmental Engineering}, \orgname{Politecnico di Milano}, \orgaddress{\street{Piazza Leonardo da Vinci, 2}, \city{Milano}, \postcode{20133}, \country{Italy}}}

\affil[3]{\orgdiv{Department of Mechanical Engineering}, \orgname{Texas A\&M University}, \orgaddress{\street{ 202 Spence St}, \city{College Station}, \postcode{ TX 77843}, \country{USA}}}

\affil[4]{\orgdiv{Instituto Universitario de Matem\'atica Pura y Aplicada}, \orgname{Universitat Polit\`ecnica de Val\`encia}, \orgaddress{\street{Camino de Vera s/n}, \city{Val\`encia}, \postcode{46022}, \country{Spain}}}

\abstract{
Diatom frustules have long fascinated scientists for the extraordinary beauty, diversity, and functionality of their intricate silica architectures.
Yet, the spatial organization of these structures has so far been primarily described in terms of morphology, symmetry, and crystallographic order, leaving common statistical properties across distinct diatom architectures largely unexplored.
By analyzing diatom frustules through the lens of hyperuniformity for the first time, we reveal a striking commonality across their remarkable diversity: all analyzed genera exhibit signatures of suppressed long-wavelength density fluctuations.
Specifically, we find that these architectures range from strongly ordered Class I to disordered Class III hyperuniform systems.
We show that local order, spatial correlations, and long-range fluctuation suppression can vary partially independently, giving rise to a continuous spectrum of multiscale architectures.
 Our multiscale approach reveals that hyperuniformity can be reliably characterized in finite biological structures, beyond what conventional asymptotic diagnostics can resolve.
Together, these results prove that hyperuniformity provides a unifying statistical framework for describing diatom diversity beyond conventional classifications of structural order.}

\keywords{Diatom $|$ Hyperuniformity $|$ Statistical physics $|$ Taxonomy $|$ Biomineralization}

\maketitle

\section*{Introduction}
Biological evolution generates an extraordinary diversity of forms, organizing matter into complex spatial architectures that accommodate multiple functional demands across scales \cite{meyers2008biological,vignolini2012pointillist,wegst2015bioinspired,nepal2023hierarchically}.
From avian photoreceptor mosaics~\cite{jiao_avian_2014} to leaf trichomes\cite{ballestero2025emergence}, natural structures often exhibit forms of order that are neither simply crystalline nor disordered, challenging the conventional binary classification of structural order. 

Diatoms, unicellular micro-algae that produce intricate patterns of silica walls, provide an exceptional natural laboratory for exploring spatial organization in biological architectures.
The estimated diversity of hundreds of thousands species of diatoms~\cite{mann2013inordinate,malviya2016global}, distributed among multiple taxonomic groups, is accompanied by an extraordinary variety of frustule architectures that span diverse shapes, symmetries, and pore organizations~\cite{round1990introduction, kroger2008diatoms, gordon2009diatoms, losic2009diatom,alverson2025phylogenomics}.
Beyond their striking morphology, frustules are remarkably multifunctional \cite{de2017diatom,musenich2025revealing}, simultaneously providing mechanical protection~\cite{hamm2003architecture, aitken2016microstructure}, regulating buoyancy~\cite{smetacek1985role, gemmell2016dynamic, kemp2018case}, and manipulating light for photosynthesis~\cite{fuhrmann2004photonic,stefano2007lensless, romann_wavelength_2015,de2016light}.

The formation of these architectures has been the subject of extensive research on diatom biomineralization \cite{poulsen2003biosilica,hildebrand2008diatoms,kroger2008diatoms,sumper2008silica}, revealing that pore formation is under tight genetic and molecular control \cite{kroger2002self,scheffel2011nanopatterned,gorlich2019control}. Studies have identified protein-mediated pore formation~\cite{heintze2022molecular}, phase-separated frustrated domains~\cite{feofilova2022geometrical}, and branching silica rib formation~\cite{babenko2024branching} as morphogenetic mechanisms underlying frustule patterning.
Against this background, the recurring presence of pore patterns throughout the remarkable diversity of diatom frustules raises the  fundamental question: do diatom frustules share common statistical  principles of spatial organization that transcend their specific geometries?

To date, existing studies of frustule architectures have focused primarily on morphological descriptors, symmetry classifications, and measures of conventional crystalline periodicity.
Losic et al.~\cite{losic2009diatom} interpreted frustule architectures as evolutionarily optimized multifunctional designs, emphasizing how variations in pore geometry and arrangement contribute to distinct mechanical, optical, transport and sensing capabilities while inspiring the development of advanced materials.
Goessling et al.~\cite{goessling2020highly} demonstrated that the girdle bands of the centric diatom Coscinodiscus granii form highly reproducible slab photonic crystals, with lattice periods precisely preserved across different cell culture strains despite variability in surface pore diameter, and showed that this preserved periodicity, rather than pore size, governs the resulting optical band gap.
Ashworth et al.~\cite{ashworth2025adaptive} extended this analysis to several hundred diatom species spanning major taxonomic groups, using Fast Fourier Transform (FFT) analysis of girdle band micrographs to identify slab-photonic-crystal lattices and map their occurrence onto a phylogenetic framework. Their results indicate that square lattice types emerged early within centric diatoms from quasi-ordered templates and subsequently diversified into hexagonal lattices, suggesting an evolutionary trajectory from quasi-order toward crystalline order or, in some lineages, toward disorder.

Collectively, these studies have established geometry and crystallographic order as the prevailing frameworks for describing frustule architectures,  typically  classified as hexagonal, quasi-crystalline, or disordered based on visual inspection of their direct-space organization or FFT analysis~\cite{ashworth2025adaptive, goessling2020highly}.
Although informative, such classifications do not provide a quantitative framework for characterizing spatial order across the continuous spectrum of architectures observed among diatom genera, nor do they reveal whether common forms of statistical properties persist across morphologically distinct species.
Addressing this gap requires moving beyond purely descriptive classifications  toward quantitative descriptors of spatial organization.

\begin{figure*}[t]
\centering
\includegraphics[width=1\textwidth]{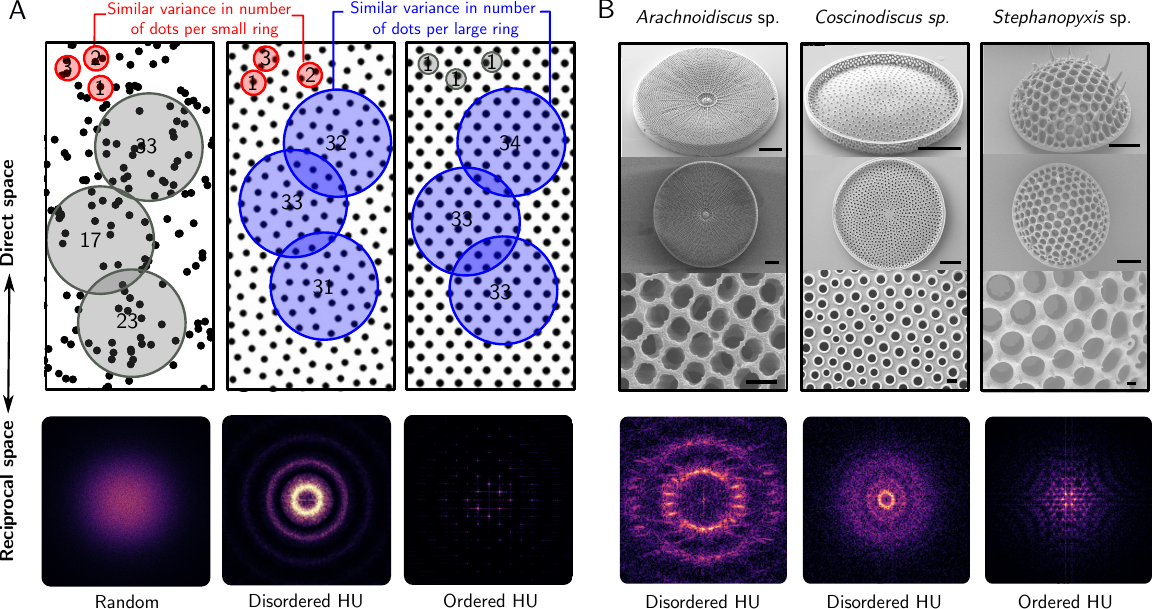}
\caption{
\textbf{Hyperuniformity as a framework for describing spatial organization in diatom frustules.}
(a) Schematic representation of random, disordered hyperuniform, and periodic point patterns in real space (top), together with their corresponding reciprocal-space representations (bottom).
Observation windows illustrate the evolution of density fluctuations across different forms of spatial organization.
In random patterns, density fluctuations persist as the observation scale increases, whereas in disordered hyperuniform systems they become strongly suppressed at large scales.
Periodic crystals exhibit strong suppression of density fluctuations imposed by periodic order.
These differences are reflected in reciprocal space, where hyperuniform systems exhibit vanishing scattering intensity as the wavevector approaches zero, while periodic crystals additionally display discrete Bragg peaks.
(b) Scanning electron microscopy (SEM) images of representative diatom frustules (\textit{Arachnodiscus} sp., \textit{Coscinodiscus} sp., \textit{Stephanopyxis} sp.) illustrating the remarkable diversity of pore architectures found across diatom genera, from apparently disordered arrangements to nearly crystalline lattices. Scale bars, \SI{20}{\micro\meter} (top and middle rows) and \SI{2}{\micro\meter} (bottom row). Despite their distinct morphologies, these architectures exhibit common signatures of hyperuniform spatial organization, suggesting that hyperuniformity provides a unifying statistical framework for describing pore organization across diatom diversity.
}
\label{Fig1}
\end{figure*}

A natural framework for quantifying such statistical spatial organization is hyperuniformity (HU).
Hyperuniformity describes a specific class of spatial organizations characterized by strongly suppressed large-scale density fluctuations~\cite{torquato_local_2003, torquato_hyperuniformity_2016, torquato_hyperuniform_2018}.
Figure~\ref{Fig1}a illustrates how hyperuniform systems span a continuum that ranges from completely disordered arrangements to perfectly ordered crystals.
In a random point pattern (left panel), both small (blue circles) and large (gray circles) observation windows intercept significantly different numbers of structural elements depending on the selected probing window, indicating persistent density fluctuations across spatial scales.
As spatial correlations develop (central panel), these fluctuations are progressively suppressed: although small observation windows (blue circles) still probe different local environments, sufficiently large windows (green circles) contain nearly the same number of structural elements regardless of their location.
In the limiting case of a perfectly periodic crystal (right panel), both small and large observation windows contain nearly the same number of structural elements regardless of their position, reflecting the strong suppression of density fluctuations imposed by periodic order.
The differences in density fluctuations illustrated in real space are directly reflected in the corresponding reciprocal-space representations (bottom panels of Fig.~\ref{Fig1}).
Random patterns exhibit a finite scattering intensity near the origin, whereas disordered hyperuniform systems suppress long-wavelength fluctuations, leading to a vanishing scattering intensity as the wavevector approaches zero.
In the limiting case of a periodic crystal, this suppression is accompanied by discrete Bragg peaks that reflect periodic order.

Here, we characterize diatom frustules through the lens of hyperuniformity for the first time, employing complementary descriptors in both direct and reciprocal space.
Specifically, we investigated 21 representative diatom genera that span a broad diversity of frustule morphologies and spatial organizations.
By combining local, intermediate, and long-range descriptors, we establish a multiscale framework for quantifying spatial order across these distinct architectures.
Our results reveal that, despite their remarkable morphological diversity, diatom frustules occupy a continuous hyperuniform morphospace.
This suggests that hyperuniformity may act as a unifying principle of spatial organization in diatom architectures.
More broadly, these findings reveal a previously hidden level of statistical organization beneath morphological diversity, which can emerge as a common principle for understanding spatial organization across otherwise distinct diatom architectures and biological systems, in general.

\section*{Results}
\subsection*{Diatom frustules exhibit multiscale hyperuniform organization}
To quantitatively characterize diatom frustules through the lens of hyperuniformity, we employ complementary descriptors in direct and reciprocal space.
For a statistically homogeneous two-phase medium, hyperuniformity can be expressed in direct space through local volume-fraction variance $\sigma_v^2(R)$, measured within an observation window of volume $v(R)$ and characteristic size $R$.
In generic random media, $\sigma_v^2(R)$ decays as $R^{-d}$ in \textit{d}-dimensional Euclidean space $\mathbb{R}^d$,i.e., as the inverse of the window volume, while hyperuniform systems exhibit an asymptotically faster suppression of large-scale fluctuations~\cite{zachary_hyperuniformity_2009, torquato_hyperuniform_2018, kim_characterizing_2021}.
This defining property can be written as
\begin{equation}
\lim_{v(R) \to \infty} v(R) \sigma^2_v(R) = 0.
\label{eq:variance_HU_def}
\end{equation}
In crystalline systems, this suppression is particularly strong, with $\sigma_v^2(R) \sim R^{-\beta}$ and $\beta > d$.

In reciprocal space, hyperuniformity is characterized by the spectral density $\Psi(\mathbf{q})$, defined as the Fourier transform of the autocovariance function $\psi(\mathbf{r})$,
\begin{equation}
\Psi(\mathbf{q}) = \int_{\mathbb{R}^d} \psi(\mathbf{r}) \, e^{-i \mathbf{q}\cdot \mathbf{r}} \, d\mathbf{r},
\end{equation}
which satisfies the condition
\begin{equation}
\lim_{|\mathbf{q}| \to 0} \Psi(\mathbf{q}) = 0,
\end{equation}
i.e., a vanishing spectral density at low wavenumbers $q = |\mathbf{q}|$~\cite{philcox_disordered_2023, torquato_hyperuniformity_2016, zachary_hyperuniformity_2009}. 

This direct / reciprocal equivalence provides the theoretical basis for our analysis.
Figure~\ref{Fig1}b shows two representative diatom frustules exhibiting distinct forms of hyperuniform spatial organization.
The first displays an apparently disordered pore arrangement, whose spectral density exhibits an isotropic ring of scattering maxima with suppressed intensity at low $q$, consistent with disordered hyperuniformity (see Fig.~\ref{Fig1}a).
In contrast, the second exhibits a nearly crystalline pore lattice, whose spectral density is characterized by anisotropic Bragg-like peaks reflecting coherent lattice-like order.

To further characterize these distinct forms of spatial organization quantitatively, we employ complementary metrics that probe different length scales.
Specifically, high-resolution SEM images of diatoms were cropped to selected regions of the frustule valves, binarized to isolate the pore structures, and converted into binary-phase and point-pattern representations for subsequent quantitative analysis.
A comparative multiscale characterization is then performed on 21 representative genera of diatoms.
For each genus, the extracted pore patterns are characterized using (i) the bond-orientational order parameter $\psi_6$ and its mean value $\langle\psi_6\rangle$, (ii) the real-space autocovariance function $\psi(\mathbf{r})$ and the associated persistence metric $\gamma$, and (iii) the spectral density $\Psi(\mathbf{q})$.
Together these descriptors allow to probe spatial organization across local, intermediate, and large length scales.

\begin{figure}[t!]
\centering
\includegraphics[width=0.5\textwidth]{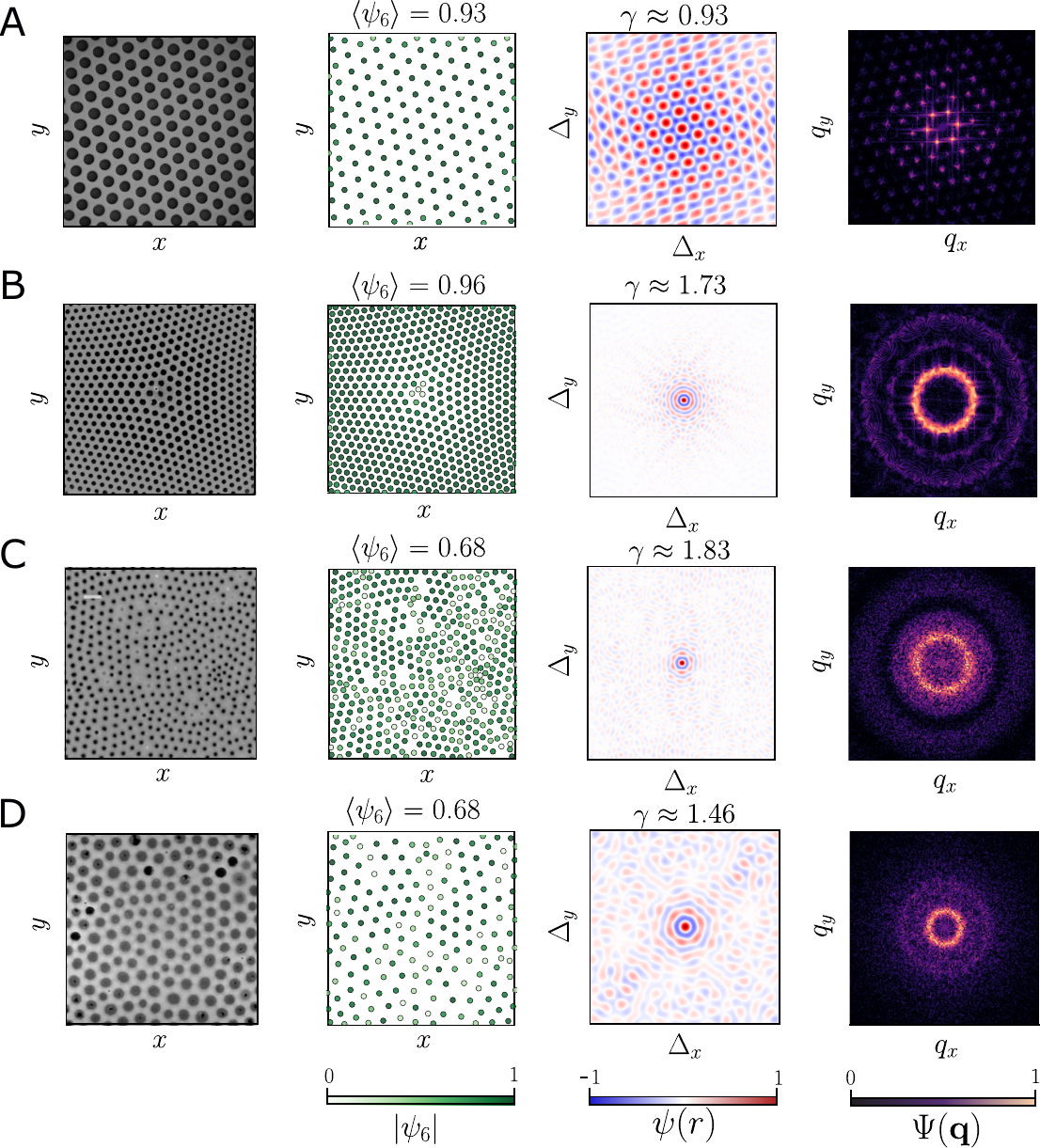}
\caption{
\textbf{Multiscale structural characterization of representative diatom frustules.}
Rows correspond to four representative diatom genera: \textit{Roperia} (A), \textit{Planktoniella} (B), \textit{Porosira} (C), and \textit{Azpeitia} (D).
Columns show, from left to right, SEM images of the complete frustule valves; pore patterns extracted from selected regions for quantitative analysis; the local bond-orientational order parameter $\psi_6$, together with its mean value $\langle\psi_6\rangle$; the real-space autocovariance function $\psi(\mathbf{r})$, together with the associated persistence metric $\gamma$; and the corresponding spectral density $\Psi(\mathbf{q})$.
}
\label{Fig2}
\end{figure}

Figure~\ref{Fig2} presents SEM images of the frustule valves of four representative diatom genera, namely Roperia, Planktoniella, Porosira, and Azpeitia together with their multiscale structural characterization.

The first column shows the complete frustule valves, illustrating the morphological diversity of the selected genera.
Visual inspection already reveals substantial differences in pore organization: Roperia and Planktoniella exhibit highly ordered, lattice-like arrangements, whereas Porosira and Azpeitia display more irregular architectures with no apparent long-range periodicity.
The second column presents the pore patterns extracted from the selected regions and used for the subsequent analysis.

To quantify the local organization, the third column reports the bond orientation order parameter $\psi_6 \in [0,1]$, which measures the degree of hexagonal symmetry around each pore.
Roperia and Planktoniella exhibit predominantly high $\psi_6$ values, with $\langle\psi_6\rangle$ close to unity, reflecting their pronounced local hexagonal order, while the greater local heterogeneity observed in Porosira and Azpeitia results in lower mean values ($\langle\psi_6\rangle \approx 0.68$).

To investigate spatial organization beyond the local scale, the fourth column presents the real-space autocovariance function $\psi(\mathbf{r})$.
The four genera exhibit markedly different correlation patterns, with oscillations persisting over different spatial ranges and decaying at different rates.
To quantify these differences, we introduce the autocovariance persistence metric $\gamma$, which characterizes the decay of the oscillatory envelope of $\psi(\mathbf{r})$ beyond the nearest-neighbor scale and within the finite observation window.

Finally, the fifth column shows the corresponding spectral density $\Psi(\mathbf{q})$, providing a complementary description of spatial organization in reciprocal space.
The four genera exhibit distinct spectral signatures, ranging from anisotropic Bragg-like peaks to ring-like or more diffuse scattering patterns, reflecting their different degrees of structural order.
Despite these differences, all four genera exhibit suppressed spectral density at low $q$, consistent with the interpretation of a hyperuniform spatial organization.

\subsection*{Distinct hyperuniform scaling regimes}
The suppression of spectral density observed in Fig.~\ref{Fig2} reveals a common signature of hyperuniformity in structurally distinct diatom frustules.
To further quantify this behavior, we examine the suppression of long-wavelength fluctuations in the low-$q$ regime through the scaling exponent $\alpha$, defined in reciprocal space as
\begin{equation}
\Psi(|\mathbf{q}|) \sim |\mathbf{q}|^\alpha,
\label{eq:alpha_scaling}
\end{equation}
where $\alpha$ is the log--log slope of the radial spectral density.
The positive values of $\alpha$ identify hyperuniform behavior, with its magnitude distinguishing three asymptotic regimes: Class~I for $\alpha>1$, Class~II for $\alpha=1$ and Class~III for $0<\alpha<1$~\cite{torquato_hyperuniform_2018, kim_characterizing_2021}.

The corresponding direct-space scaling of the local volume-fraction variance is
\begin{equation}
\sigma^2_{v(R \to \infty)} \sim
\begin{cases}
R^{-(d+1)}, & \alpha > 1 \quad\quad\quad \mathrm{(Class\,I)} \\
R^{-(d+1)} \ln R, & \alpha = 1 \quad\quad\quad \mathrm{(Class\,II)} \\
R^{-(d+\alpha)}, & 0 < \alpha < 1 \quad \mathrm{(Class\,III)}
\end{cases}.
\label{eq:eq_HU_classes}
\end{equation}

Equation~\ref{eq:eq_HU_classes} allows us to independently quantify the suppression of density fluctuations through the scaling descriptor $\sigma_v^2(R)\sim R^{-\beta}$, where $\beta$ describes the decay of the local volume-fraction variance with increasing observation-window size $R$.
For $0<\alpha<1$, the two exponents are related by $\beta=d+\alpha$ (with $d=2$ here), for $\alpha\geq1$, $\beta$ reaches the limiting value $\beta=d+1$, with a logarithmic correction at $\alpha=1$.
Thus, $\beta$ provides an independent direct-space measure of the suppression of large-scale density fluctuations, complementing the reciprocal-space characterization.
However, because $\beta$ saturates at $d+1$ for $\alpha>1$, direct-space analysis alone cannot resolve variations among Class~I systems, for which reciprocal-space measurements remain essential.

\begin{figure*}[t!]
\centering
\includegraphics[width=1\textwidth]{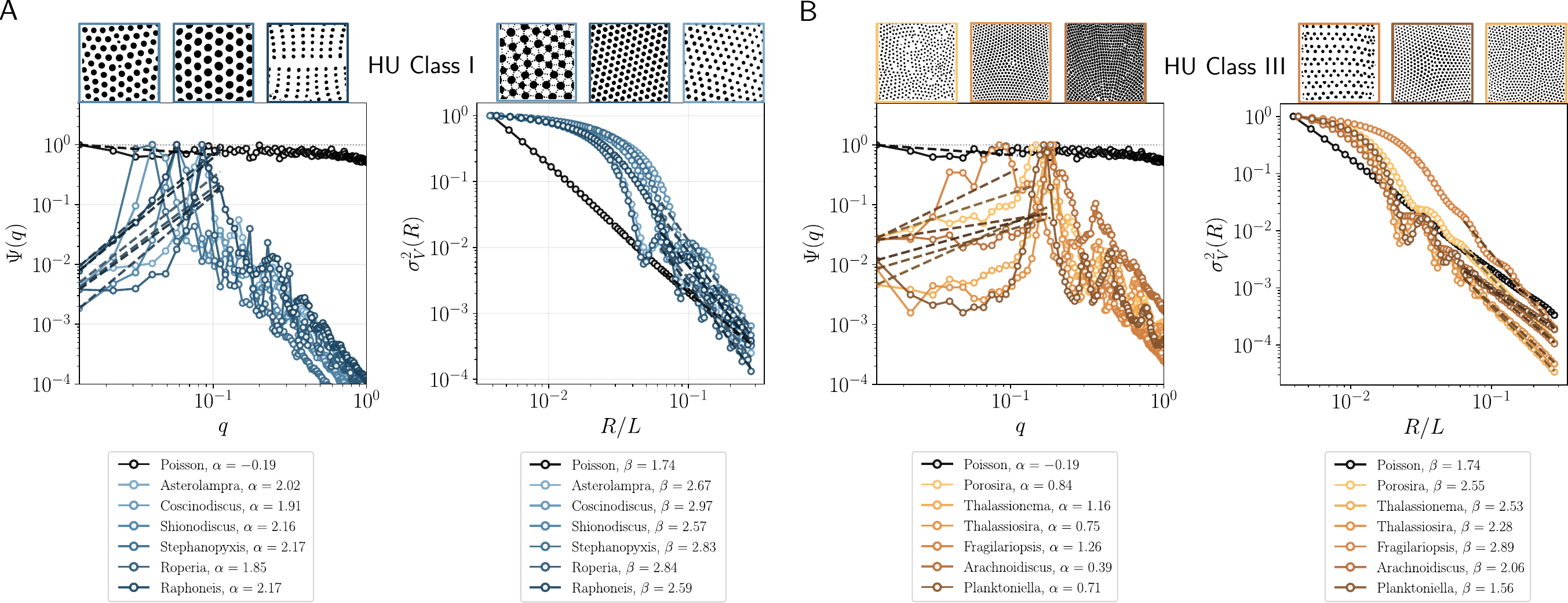}
\caption{\textbf{Reciprocal- and direct-space scaling behavior of representative diatom genera across distinct hyperuniform regimes.}
Radially averaged spectral density $\Psi(q)$ (left panels) and local volume-fraction variance $\sigma_v^2(R)$ (right panels) for representative diatom genera spanning different regimes of hyperuniformity.
(A) Genera exhibiting Class~I hyperuniformity, characterized by reciprocal-space scaling exponents $\alpha>1$ and direct-space exponents approaching the limiting decay rate $\beta=d+1$.
(B) Genera predominantly exhibiting Class~III hyperuniformity, characterized by $0<\alpha<1$ and intermediate direct-space scaling exponents $d<\beta<d+1$.
Dashed lines indicate the linear regressions used to estimate the reciprocal-space exponent $\alpha$ in the low-$q$ regime and the direct-space exponent $\beta$ in the large-$R$ regime.
Insets show representative SEM crops of the corresponding frustule pore architectures.
A synthetic Poissonian pattern, yielding $\alpha\approx0$ and $\beta\approx d$, is included as a reference for non-hyperuniform behavior and shown as a dashed black line.}
\label{Fig3}
\end{figure*}

Figure~\ref{Fig3} reports the reciprocal- and direct-space scaling behavior of representative diatom genera spanning different hyperuniform regimes, quantified through (left panels) the radially averaged spectral density $\Psi(q)$ and (right panels) local volume-fraction variance $\sigma_v^2(R)$, respectively.
The analyzed genera separate into distinct scaling regimes. All genera shown in Fig.~\ref{Fig3}A exhibit $\alpha>1$, corresponding to Class~I hyperuniformity, whereas those shown in Fig.~\ref{Fig3}B predominantly exhibit $0<\alpha<1$, corresponding to Class~III hyperuniformity, with two genera displaying $\alpha$ values slightly above unity.
Despite the rather different pore architectures, illustrated by the SEM crops in the insets, all analyzed genera exhibit $\alpha>0$, confirming that the suppression of spectral density observed in Fig.~\ref{Fig2} persists asymptotically toward the low-$q$ regime in all the analyzed genera.
Looking at the corresponding local volume-fraction variances (right panels of Figs.~\ref{Fig3}A,B), a general faster decay than the Poissonian reference is observed, with $\beta>d$ for all genera except Planktoniella.
Genera such as Roperia and Coscinodiscus approach the Class~I limit $\beta=d+1$ exhibiting $\beta = 2.97$ and $\beta = 2.84$, respectively, whereas Porosira and Thalassionema exhibits intermediate values of $d<\beta<d+1$, namely $\beta = 2.55$ and $\beta = 2.53$.
A synthetic Poissonian pattern, yielding $\alpha\approx0$ and $\beta\approx d$, is included as a reference for non-hyperuniform behavior and reported as the dashed black line.

The exponent $\alpha$ was estimated through an adaptive linear regression of the low-$q$ spectral density, extending from the first radial bin to the onset of the first dominant scattering feature.
The dashed lines in Fig.~\ref{Fig3} indicate the corresponding regressions.
This procedure captures the progressive low-$q$ increase of $\Psi(q)$ while limiting the influence of residual DC contamination near the origin and avoiding overestimation from a direct origin-to-peak fit.
The adaptive regression is particularly relevant for finite samples with limited low-$q$ resolution, such as Planktoniella and Fragilariopsis, for which multiple images are not available for ensemble averaging.
The exponent $\beta$ was independently estimated from a linear regression of the large-$R$ decay of $\sigma_v^2(R)$.
Superdisk averaging was used to reduce biases associated with the different degrees of anisotropy exhibited by frustule architectures~\cite{kim_characterizing_2021}.

Despite the general good agreement between reciprocal- and direct-space descriptors, the estimated values of $\alpha$ and $\beta$ do not always satisfy the ideal asymptotic relation.
Unlike infinite theoretical systems, the analyzed frustule images are finite, heterogeneous, and, in some cases, anisotropic, while segmentation and projection effects further limit access to the low-$q$ and large-$R$ regimes in which asymptotic scaling is expected to emerge.
These limitations are particularly evident for Planktoniella, which exhibits $\alpha>0$ but $\beta<d$.
However, recent work has shown that appropriately defined local and intermediate-scale metrics can nevertheless provide robust signatures of hyperuniformity in confined systems of moderate size~\cite{kim_characterizing_2021}.
Consequently, rather than treating $\alpha$ and $\beta$ as interchangeable quantities constrained to satisfy their ideal asymptotic relation, we interpret them here as complementary descriptors of long-range spatial organization in finite biological structures.

\subsection*{Intermediate-scale correlations}
Figure~\ref{Fig3} revealed substantial variability in hyperuniform scaling across diatom genera, ranging from steep spectral scaling and persistent real-space decay associated with more crystalline Class~I architectures to weaker but still positive low-$q$ scaling characteristic of more disordered Class~III systems.
To determine whether this variability is also reflected at intermediate spatial scales, we quantify the persistence of structural correlations through the decay exponent $\gamma$ of the autocovariance envelope.

\begin{figure}[t!]
\centering
\includegraphics[width=0.5\textwidth]{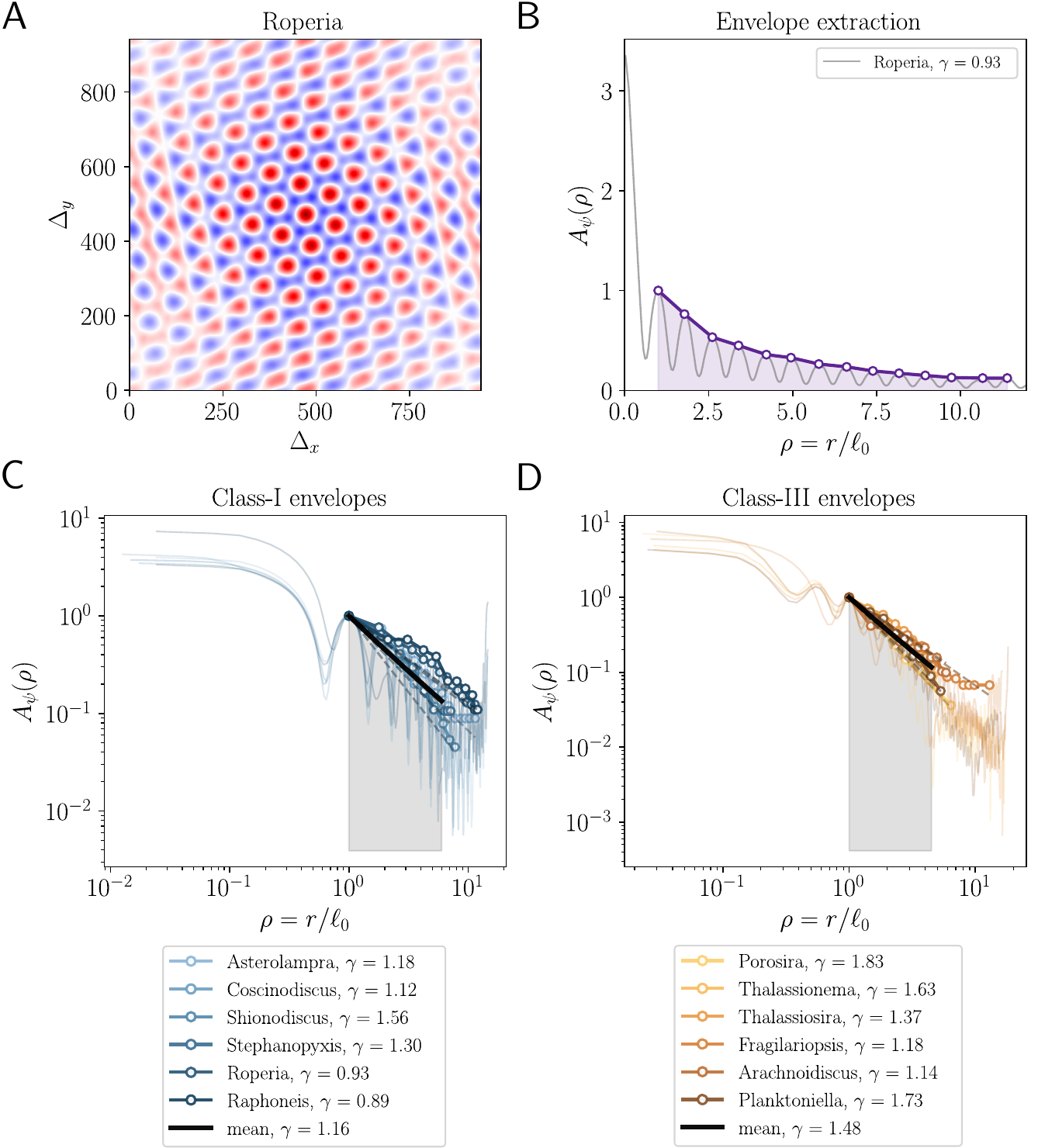}
\caption{\textbf{Intermediate-scale persistence of spatial correlations across diatom genera.}
(A) Spatial autocovariance field $\psi(\mathbf{r})$ of Roperia, shown as a representative example.
(B) Corresponding autocovariance envelope $A_\psi(\rho)$ as a function of the normalized distance $\rho=r/\ell_0$, where $\ell_0$ is the characteristic pore spacing.
The correlation persistence exponent $\gamma$ is estimated from the scaling $A_\psi(\rho)\sim\rho^{-\gamma}$.
(C,D) Autocovariance-envelope decay for representative diatom genera exhibiting predominantly Class~I and Class~III hyperuniformity, respectively.
Lower values of $\gamma$ indicate slower decay and more persistent spatial correlations, whereas higher values indicate faster decorrelation.}
\label{Fig4}
\end{figure}

Figures~\ref{Fig4}A,B illustrate this analysis using Roperia as a representative example.
Starting from its spatial autocovariance $\psi(r)$ (Fig.~\ref{Fig4}A), $\gamma$ is extracted from the decay of its oscillation envelope with normalized distance $\rho=r/\ell_0$ (Fig.~\ref{Fig4}B).
Low values of $\gamma$ correspond to slowly decaying oscillations and persistent spatial correlations, whereas higher values indicate faster decorrelation.

Extending this analysis across genera reveals substantial variability in the decay of spatial correlations (Figs.~\ref{Fig4}C,D).
Specifically, genera associated with more ordered Class~I architectures generally exhibit slower correlation decay and lower values of $\gamma$, while more disordered, predominantly Class~III hyperuniform architectures tend to exhibit faster decay and higher values of $\gamma$.
Importantly, the values of $\gamma$ vary continuously across genera rather than separating into sharply distinct groups, indicating that the diatom frustules span a continuum of spatial organizations between strongly correlated and increasingly disordered hyperuniform architectures.

Together, the local bond-orientational order parameter $\langle\psi_6\rangle$, the intermediate-range exponent $\gamma$, and the long-range scaling exponents $\alpha$ and $\beta$ provide a multiscale description of spatial organization (see Table~\ref{tab:HU_metrics}): $\langle\psi_6\rangle$ quantifies local symmetry, $\gamma$ measures the persistence of spatial correlations, and $\alpha$ and $\beta$ characterize the suppression of fluctuations at the largest accessible scales.
Their combined analysis is particularly relevant for finite biological structures, for which the asymptotic limits underlying conventional hyperuniformity diagnostics cannot always be fully accessed.
By integrating local, intermediate-, and long-range descriptors, this multiscale framework goes beyond discrete hyperuniform classification and reveals a continuum of spatial organizations across diatom genera.

\begin{table}[t]
\centering
\caption{Multi-scale descriptors used to characterize hyperuniform spatial organization in diatom frustules.}
\label{tab:multiscale_metrics}
\footnotesize
\begin{tabular}{p{0.12\columnwidth} p{0.10\columnwidth} p{0.63\columnwidth}}
\hline
\textbf{Scale}  & \textbf{Symbol}         & \textbf{Physical meaning} \\
\hline
Local   & $\langle \psi_6 \rangle$ & Hexagonal symmetry / short range order. \\
Medium  & $\gamma$                    & Persistence of oscillatory autocorrelations in $\psi(r)$. \\
Large   & $\alpha$                 & Low-$q$ suppression of density fluctuations in $\Psi(\mathbf{q})$. \\
Large   & $\beta$                  & Large-$R$ suppression of density variance in $\sigma_v^2(R)$ \\
\hline
\end{tabular}
\label{tab:HU_metrics}
\end{table}

\subsection*{Hyperuniformity spans diatom taxonomic diversity}
\begin{figure*}[t!]
\centering
\includegraphics[width=\textwidth]{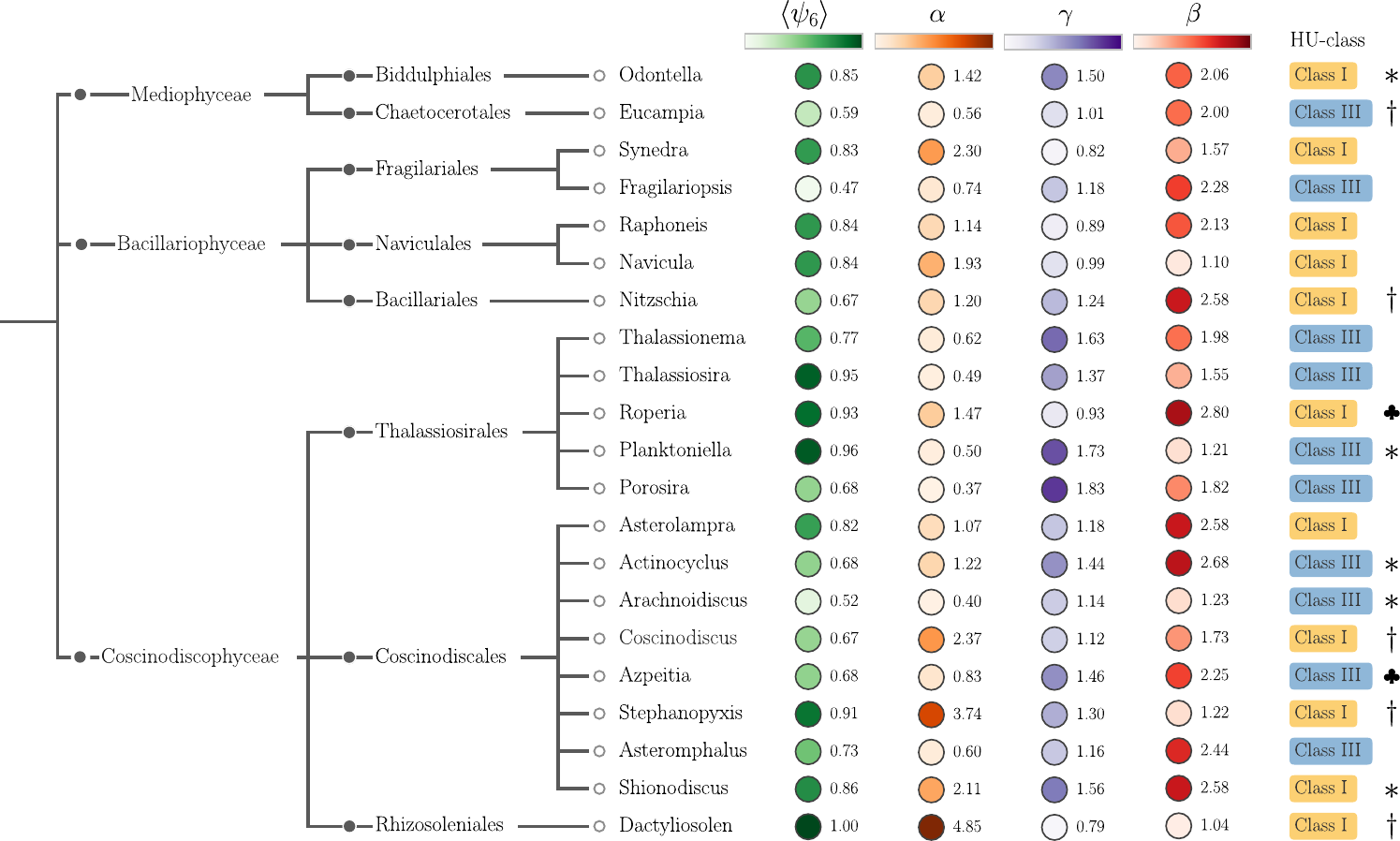}
\caption{\textbf{Multiscale hyperuniform organization across diatom taxonomic diversity.}
Taxonomic organization of the 21 analyzed diatom genera, spanning three major classes, together with their local bond-orientational order $\langle|\psi_6|\rangle$, intermediate-scale correlation persistence exponent $\gamma$, long-range scaling exponents $\alpha$ and $\beta$, and corresponding hyperuniform class.
Asterisks indicate genera for which the multiscale descriptors do not converge toward a unique hyperuniform class, whereas daggers identify cases in which finite image dimensions limit reliable access to the asymptotic scaling regimes.
Club symbols identify representative genera highlighted in the main text.
Genus-specific real- and reciprocal-space analyses, complete descriptor values, and classification criteria are provided in the SI.}
\label{Fig5}
\end{figure*}

Integrating the multiscale descriptors across all 21 analyzed genera reveals how hyperuniform spatial organization is distributed throughout diatom taxonomic diversity (Fig.~\ref{Fig5}).
The structural map combines local bond-orientational order $\langle|\psi_6|\rangle$, intermediate-scale correlation persistence $\gamma$, and long-range scaling exponents $\alpha$ and $\beta$ within the taxonomic organization of the analyzed genera.
This representation provides a comparative map of spatial organization across taxa, allowing structural similarities and differences to be examined across multiple scales.
The complete set of $\langle|\psi_6|\rangle$, $\gamma$, $\alpha$, and $\beta$ values for all analyzed genera is provided in the SI.

Hyperuniformity is widespread across the analyzed genera, with approximately half exhibiting predominantly Class~I behavior and half predominantly Class~III behavior.
Class~I architectures generally combine stronger local orientational order, more persistent spatial correlations, and steeper low-$q$ spectral scaling.
Class~III architectures, in contrast, exhibit weaker local order, faster decorrelation, and shallower but still positive low-$q$ scaling.
Both regimes occur across distinct taxonomic groups rather than being confined to individual classes or orders.
For example, Class~I architectures occur in genera as morphologically and taxonomically distinct as Roperia, Raphoneis, and Coscinodiscus, whereas Class~III behavior is observed in Porosira, Thalassiosira, and Azpeitia.

The distribution of multiscale descriptors also reveals that the boundaries between hyperuniform regimes are not always sharp.
Several genera exhibit combinations of local, intermediate, and long-range metrics that do not converge toward a single ideal asymptotic class, particularly when finite image dimensions limit access to the low-$q$ and large-$R$ regimes.
These intermediate cases reveal that diatom frustules populate a broad structural spectrum between strongly ordered and more disordered hyperuniform architectures.
Together, these results show that hyperuniformity is not restricted to a specific morphology or taxonomic lineage but emerges across the evolutionary diversity of diatoms through continuously varying forms of spatial organization.

Genera marked with an asterisk correspond to initially unresolved classifications requiring manual interpretation, whereas dagger symbols indicate cases where finite image dimensions likely affected parameter extraction reliability. In both cases, a HU class was assigned after a more informed study of the viability of each metric. 

\subsection*{Diatom frustules span a continuous hyperuniform morphospace}
\begin{figure*}[t!]
\centering
\includegraphics[width=0.95\textwidth]{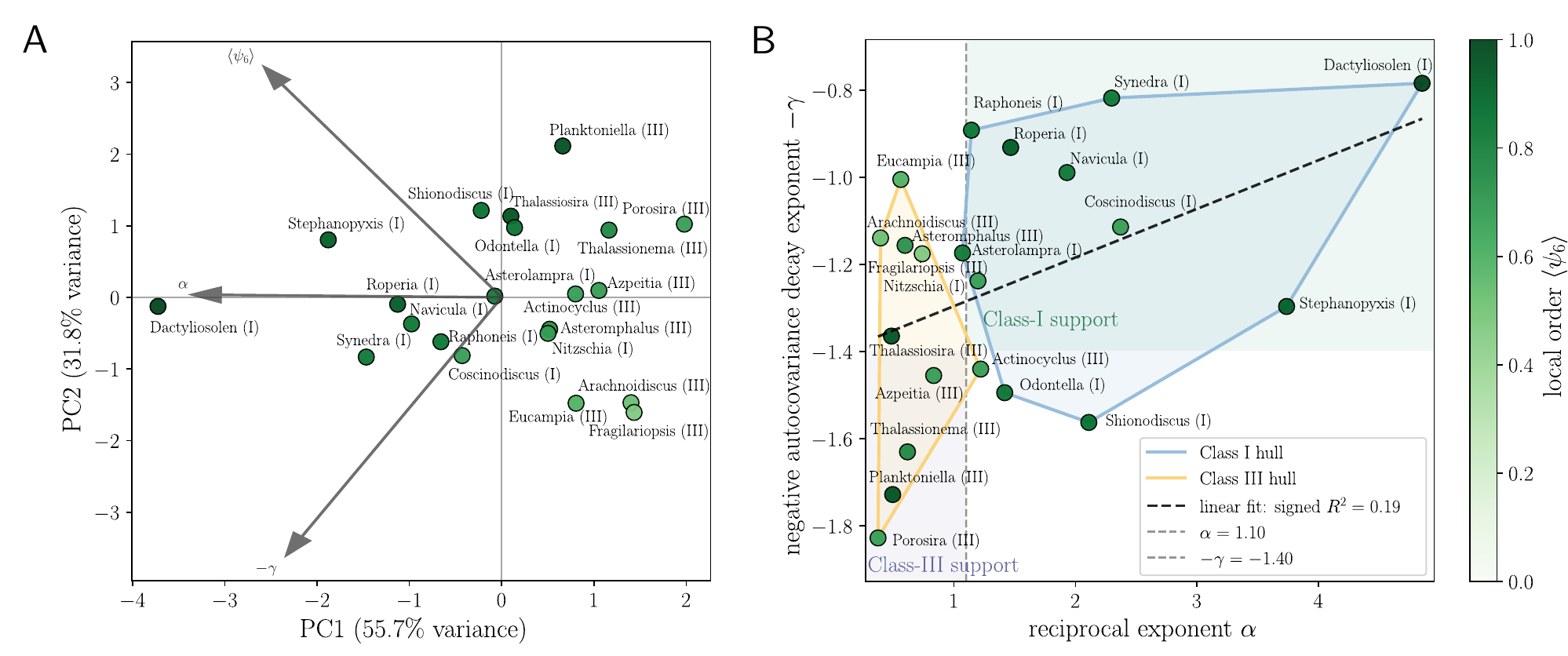}
\caption{\textbf{Multiscale descriptor-space analysis of hyperuniform organization across diatom genera.}
(A) Principal component analysis (PCA) of the standardized descriptor triplet $(\alpha,-\gamma,\langle|\psi_6|\rangle)$ for the 21 analyzed genera.
Each point represents a genus, while arrows indicate the loading directions of the three descriptors.
The first two principal components account for $87.5\%$ of the total variance.
(B) Distribution of the analyzed genera in the $(\alpha,-\gamma)$ descriptor plane.
Convex hulls indicate the regions occupied by Class~I and Class~III genera, and the dashed black line shows the linear fit between the two descriptors ($R^2=0.19$).
In both panels, point color denotes the local bond-orientational order parameter $\langle|\psi_6|\rangle$.}
\label{Fig6}
\end{figure*}

To examine how spatial organization on different scales collectively varies between diatom genera, we performed a principal component analysis (PCA) on the standardized descriptor triplet $(\alpha,-\gamma,\langle|\psi_6|\rangle)$ (Fig.~\ref{Fig6}A).
The sign convention $-\gamma$ was adopted so that larger values correspond to more persistent spatial correlations.
The exponent $\beta$ was excluded due to its theoretical redundancy to $\alpha$ in the infinite-system limit and its lower robustness in finite images, where reliable estimation is particularly sensitive to the largest accessible observation windows.
The PCA reveals how local order, intermediate-scale correlation persistence, and long-range fluctuation suppression jointly vary across hyperuniform diatom architectures.

The first two principal components account for $87.5\%$ of the total variance, indicating that most inter-genus variability can be represented within a two-dimensional descriptor space.
The first principal component (PC1, $55.7\%$ of the variance) is primarily associated with the reciprocal-space exponent $\alpha$, with additional contributions from the correlation-persistence exponent $\gamma$ and local bond-orientational order $\langle|\psi_6|\rangle$.
Genera with steeper low-$q$ spectral scaling and more persistent spatial correlations preferentially extend toward negative PC1 values, whereas genera with weaker spectral suppression and faster decorrelation predominantly occupy positive PC1 values.
Accordingly, PC1 captures the dominant variation between the more strongly correlated Class~I architectures and the more disordered Class~III architectures, although the two regimes are not sharply separated.
The second principal component (PC2, $31.8\%$ of the variance) primarily contrasts local bond-orientational order with intermediate-scale correlation persistence, as reflected by the opposite orientations of the $\langle|\psi_6|\rangle$ and $-\gamma$ loading vectors.
This result indicates that local geometric order and the persistence of spatial correlations represent distinct, only partially coupled dimensions of frustule organization.
Consequently, similar levels of local hexagonal order can coexist with different degrees of correlation persistence and long-wavelength fluctuation suppression, while comparable hyperuniform scaling can emerge from architectures with different local organizations.
Diatom frustules, therefore, cannot be described along a single order--disorder axis but instead exhibit partially independent structural variation across spatial scales.

The complementary descriptor-space representation in Fig.~\ref{Fig6}B further illustrates how Class~I and Class~III genera distribute within the $(\alpha,-\gamma)$ plane.
The weak correlation between the two descriptors ($R^2=0.19$) demonstrates that correlation persistence does not simply reproduce the information encoded by the reciprocal-space exponent.
Indeed, genera with comparable values of $\alpha$ can exhibit substantially different correlation persistence, while similar values of $\gamma$ coexist with markedly different spectral exponents.
Class~I and Class~III genera therefore preferentially occupy different but partially overlapping regions of the descriptor space, as highlighted by their corresponding convex hulls.
Class~I architectures span a broad range of $\alpha$ and extend toward more persistent spatial correlations, whereas Class~III architectures are predominantly concentrated at lower $\alpha$ and faster correlation decay.
The close proximity of the two regions near the formal Class~I--III threshold, together with the broad distribution of genera within each region, indicates that discrete hyperuniform classes are embedded within a richer multiscale spectrum of spatial organizations.

\section*{Discussion}
Twenty-one diatom genera have been investigated through the lens of hyperuniformity for the first time, revealing a common statistical framework through which otherwise distinct diatom architectures can be quantitatively compared.
Our results showed that all analyzed genera exhibit signatures of suppressed long-wavelength density fluctuations, spanning architectures from (i) strongly ordered Class~I hyperuniform systems to (ii) disordered Class~III hyperuniform systems.
These findings uncovered a level of spatial organization that had remained hidden so far, as previous analyses of diatom architectures have relied primarily on conventional descriptors of morphology, symmetry, and crystallographic order.

To characterize hyperuniformity reliably in these biological structures, we combined reciprocal-space spectral density $\Psi(q)$ and direct-space local volume-fraction variance $\sigma_v^2(R)$ through their respective scaling exponents $\alpha$ and $\beta$.
This complementary analysis accounted for the limited access to asymptotic regimes and the imperfect correspondence between reciprocal- and direct-space scaling in finite structures.
The consistency of the resulting classification across these descriptors demonstrated the robustness of the proposed approach, indicating that hyperuniformity can be reliably characterized in finite biological structures, even when conventional asymptotic diagnostics alone are insufficient.
Furthermore, the multiscale perspective proposed here provided a physical interpretation of the continuous structural spectrum revealed across diatom genera.
Specifically, the partial independence of local orientational order, intermediate-scale correlation persistence, and long-range fluctuation suppression showed that diatom frustules can combine these forms of spatial organization in different proportions, rather than being positioned along a simple crystalline-to-disordered axis.
As a consequence, the resulting architectures span a continuous multiscale morphospace, ranging from highly coherent pore lattices exhibiting anisotropic Bragg-like peaks to more heterogeneous structures characterized by isotropic Bragg-like rings and weaker, but still positive, low-$q$ scaling.

The recurrence of hyperuniformity across taxonomically distinct diatom genera raises fundamental questions about its biological and evolutionary origin.
This shared organizing principle, found across evolutionarily distant lineages, suggests that physical constraints on pattern formation, rather than genetic lineage alone, shape diatom architecture and that distinct developmental processes may converge toward structures sharing common statistical properties.

The nonuniform distribution of structural descriptors across taxa further indicates that evolutionary constraints may shape how local order, correlation persistence, and large-scale fluctuation suppression are combined in a given species.
Testing this hypothesis will require integrating the present approach with phylogenetic data, a step beyond the scope of the current analysis, which is based on two-dimensional SEM representations of a limited sample and offers a comparative, rather than phylogenetic, taxonomic map.
Extending it to three-dimensional structural data and explicit phylogenetic comparisons will be an important step toward establishing whether the regimes identified here reflect common developmental mechanisms or convergent evolutionary pathways.

A natural direction for future work stemming from this investigation is to establish how this statistical order translates into the physical properties of the frustule itself.
Diatoms are ubiquitous across aquatic environments, and these unicellular microalgae are responsible for producing approximately 20-25$\%$ of atmospheric oxygen~\cite{armbrust2009life,malviya2016global}, making their frustule architecture a resource of considerable scale rather than a mere structural curiosity.
Hyperuniform architectures are known, in synthetic systems, to govern the transport of waves, suppressing scattering and opening tailored band gaps for both phonons and photons.
Investigating the transport properties of diatom frustules, which share this common statistical order across genera, may reveal an entire class of self-assembled microscale systems with built-in filtering and transport properties, with enormous potential for nanotechnology applications, either directly or as biotemplates for nanofabrication at scales and geometries that remain challenging to achieve synthetically.

\section*{Materials and Methods}

\subsection*{Diatom specimens}Diatom frustules were sourced from Stefano Barone (Diatom Shop, Italy).  Specimens were provided as dried valves individually mounted on standard microscope glass slides, a preparation method designed to preserve structural integrity and facilitate direct observation of both valve-scale morphology and pore-scale architecture.  
All samples were examined in the as-received condition, without further chemical cleaning or mechanical processing.

\subsection*{Image Acquisition and Preprocessing}
We compiled a curated dataset of high-resolution scanning electron microscopy (SEM) images spanning 21 genera of marine diatoms from three taxonomic classes: Bacillariophyceae, Coscinodiscophyceae, and Mediophyceae. For 15 genera, we acquired original SEM images in-house at IEMN (Villeneuve d'Ascq, France) using a Zeiss Ultra55 scanning electron microscope. For the remaining genera, images were obtained from curated, publicly available SEM material provided by the UMR CNRS Mediterranean Institute of Oceanography. Each image was manually selected for clarity, minimal distortion, and taxonomic traceability, then cropped to isolate individual valves and binarized by adaptive thresholding (Otsu or local-mean) to segment the pore structures.

\subsection*{Point-pattern and binary-phase representations.}
From each binarized image we derived two complementary representations: (i) binary-phase (BP) images, in which pores are treated as the void phase and silica as the matrix, and (ii) point-pattern (PP) distributions, obtained from pore centroids via connected-component labeling with sub-pixel Gaussian fitting when needed. BP representations preserve pore size, connectivity, and porosity, while PP representations isolate positional correlations; both feed the descriptors below.
 
\subsection*{Hyperuniformity descriptors}
The theoretical basis for characterizing hyperuniformity in direct space (local volume-fraction variance $\sigma_v^2(R)$, Eq.~1) and reciprocal space (spectral density $\Psi(q)$, Eqs.~2--3), together with their asymptotic relation and the three hyperuniformity classes (Eqs.~4--5), is given in the main text. We describe below how each descriptor was estimated from finite SEM images.
 
\textit{Local bond-orientational order, $\langle\psi_6\rangle$.} For each point pattern we computed, per pore $j$,
\begin{equation}
\psi_6(\mathbf{x}_j) = \frac{1}{N_j}\sum_{k=1}^{N_j} \exp(i 6 \theta_{jk}),
\label{eq:psi6}
\end{equation}
where $N_j$ is the number of nearest neighbors of pore $j$ and $\theta_{jk}$ is the bond angle to neighbor $k$ relative to a fixed reference axis. $\langle\psi_6\rangle$ is the average of $|\psi_6|$ over all pores in the pattern, quantifying local hexagonal symmetry independently of large-scale fluctuation suppression.
 
\textit{Spectral density, $\Psi(q)$, and exponent $\alpha$.} $\Psi(q)$ was computed as the radially averaged Fourier transform of the binary-phase autocovariance and superdisk-averaged to reduce anisotropy biases (41). The exponent $\alpha$ (Eq.~4) was estimated via adaptive linear regression of the low-$q$ regime, extending from the first radial bin to the onset of the first dominant scattering feature, limiting the influence of residual DC contamination near the origin.
 
\textit{Local volume-fraction variance, $\sigma_v^2(R)$, and exponent $\beta$.} $\sigma_v^2(R)$ was computed directly from the BP images over increasing observation-window sizes $R$, with superdisk averaging applied for the same reason as above. $\beta$ was obtained from a linear regression of the large-$R$ decay (Eq.~5). Because $\beta$ saturates at $d+1$ for $\alpha>1$, it cannot resolve variation within Class~I and is treated as complementary to, rather than a substitute for, $\alpha$.
 
\textit{Autocovariance persistence, $\gamma$.} Local maxima $\{r_i\}$ of the radial autocovariance $\psi(r)$ were extracted and rescaled as
\begin{equation}
\rho_i = \frac{r_i}{\ell_0},
\label{eq:rho}
\end{equation}
where $\ell_0$ is the characteristic nearest-neighbor spacing. The resulting envelope $A_\psi(\rho)$ was fit as
\begin{equation}
A_\psi(\rho) \sim \rho^{-\gamma}
\label{eq:gamma}
\end{equation}
by log--log regression. Low $\gamma$ indicates slowly decaying, persistent correlations; high $\gamma$ indicates rapid decorrelation. $\gamma$ probes medium-range coherence and is distinct from the asymptotic exponents $\alpha$ and $\beta$.
 
We note that the exact asymptotic expansion linking $\sigma_v^2(R)$ to the moments of the autocovariance function (leading and sub-leading coefficients) requires a wider dynamic range in $R$ than is accessible in finite biological images; we therefore estimated $\alpha$ and $\beta$ directly from $\Psi(q)$ and $\sigma_v^2(R)$ rather than from fitted expansion coefficients (see \textit{SI Appendix} for the full expansion and a discussion of finite-size limitations).
 
\subsection*{Principal component analysis.}
We performed PCA on the standardized descriptor triplet $(\alpha, -\gamma, \langle\psi_6\rangle)$ --- the sign of $\gamma$ was flipped so that larger values consistently indicate more persistent correlations --- using singular value decomposition of the $z$-scored descriptor matrix. $\beta$ was excluded owing to its redundancy with $\alpha$ in the infinite-system limit and its lower robustness in finite images (see main text). Principal components are linear combinations of the three standardized descriptors; loadings and explained variance are reported in Fig.~6.

\section*{Author Contributions}
E.B., C.G. and M.M. designed the research.
E.B. performed the analysis.
E.B., C.G., R.K.P., V.R.-G., and M.M. interpreted the results. E.B., C.G. and M.M. wrote the manuscript with input from all authors.

\bibliography{biblio.bib}
\end{document}